# Ultra-compact Silicon Multimode Waveguide Bends Based on Special Curves for Dual Polarizations

Juanli Wang,[1,2,] Shangsen Sun,[1,2,] Runsen Zhang,[1,2] Fengchun Zhang,[1,2,*] and Ning Zhu[1,2,3,*]

[1]South China Normal University, Institute of Semiconductor Science and Technology, Nanhai Campus, Foshan, China
[2]Guangdong Engineering Technology Research Center of Low Carbon and New Energy Materials, Guangzhou, China
[3]SCNU Qingyuan Institute of Science and Technology Innovation Co., Ltd., Qingyuan, China

**Corresponding Author**

Ning Zhu*, Fengchun Zhang*

* E-mail: fengchun_zhang@126.com
* E-mail: zhuning@scnu.edu.cn

**Abstract**: The multimode waveguide bends (MWBs) with very compact sizes are the key building blocks in the applications of different mode-division multiplexing (MDM) systems. To further increase the transmission capacity, the silicon multimode waveguide bends for dual polarizations are of particular interest considering the very distinct mode behaviors under different polarizations in the silicon waveguides. Seldom silicon MWBs suitable for both polarizations have been studied. In this paper we analyze several dual-polarization-MWBs based on different bending curve functions. These special curve-based silicon MWBs have the advantages of easy fabrication and low loss compared with other structures based on the subwavelength structures such as gratings. A comparison is made between the free-form curve, Bezier curve, and Euler curve, which are used in the bending region instead of a conventional arc. The transmission spectra of the first three TE and TM modes in the silicon multimode waveguide with a core thickness of 340 nm are investigated. The simulation results indicate that in the premise of the same effective radius which is only 10 μm in this paper, the 6-mode MWB based on the free-form curve has the optimal performances, including an extremely low loss below 0.052dB and low crosstalk below -25.97dB for all six modes in the wide band of 1500-1600 nm. The MWBs based on the Bezier and Euler curve have degraded performances in terms of the loss and crosstalk. The results of this paper provide an efficient design method of the polarization insensitive silicon MWBs, which may leverage the researches for establishing complicated optical transmission systems incorporating both the MDM and polarization-division multiplexing (PDM) technology.

## 1.Introduction

With the sharply increasing demand for the bandwidth of data transmission, on-chip optical interconnect has attracted great interests for its potential to break the bandwidth-bottleneck in the optical communication. In order to increase the transmission capacity of optical network, various multiplexing technologies have been proposed as promising solutions, including the wavelength-division multiplexing(WDM)[1], polarization-division multiplexing (PDM)[2] and mode-division multiplexing (MDM)[3], etc. Among these technologies mentioned above, WDM and PDM have been widely utilized, while MDM is currently becoming more and more attractive because of its ability to further increase the link capacity significantly which has almost been pushed forward to its limit by the commercial WDM systems.

In MDM system, multiple guided modes of each wavelength in the multimode waveguide can be simultaneously utilized to carry the independent signals, thus efficiently enhancing the capacity for each wavelength channel. To realize an on-chip MDM system, various devices have been proposed, including the multimode waveguide crossing[4], mode (de)multiplexers[5, 6], multi-mode switches[7, 8], and multi-mode waveguide bends [9-11]. As an essential component, the waveguide bends are used to change the propagation direction of the guided light. As a result, a sharp waveguide bend is of great importance to achieve high density photonic integrated circuits[9-17].

However, the multi-mode waveguide bends in the silicon-on-insulator (SOI) nanophotonic circuits suffer serious problems of the transmission efficiencies and inter-mode crosstalk compared with those single mode bends. For a single-mode silicon waveguide bend, low excess loss is easily achieved with a small bending radius because of the high index contrast[12], and no inter-mode crosstalk is caused since the single-mode condition is maintained all the time. But when it comes to a sharp multi-mode waveguide bend, the inter-mode coupling between different eigen-modes occurs due to the mode field mismatch at the junction of the straight waveguide and the bending part. The silicon multi-mode waveguide usually requires a large bending radius which is at least two orders of magnitude larger than that of the single mode waveguide to relieve this mode mismatch. To tackle this problem, curve engineering and mode engineering are often applied to reduce the bending radius. The first one is to use special curves to design the waveguide bends[14-17], among which the Bezier curves and Euler curves are most widely used. These types of bends use curves with gradually changing radii to reduce the mode mismatch at the interface between the straight and bent waveguide. Multimode waveguide bends of this kind supporting three to four modes with the effective bending radii of 20-45 µm are proposed[14, 15], with the theory crosstalk level of around -23 to -25dB. However, a four-mode MWB with a bending radius smaller than 20 µm with acceptable performance is usually difficult to achieve with this kind of curve functions. The other method utilizes the mode converters, by which the modes in the straight waveguide will be converted into the distorted modes in the bent waveguide, or vice versa, so as to reduce the inter-mode coupling[10, 11, 18]. The smallest radius of 10 µm is realized for a waveguide bend with top subwavelength gratings (SWGs) supporting three modes[11]. Apart from these two mechanisms mentioned above, the vertical multi-mode waveguide is also used to acquire small bending radius since it is single-mode in the lateral dimension[19], being less popular due to the reason that it has a larger aspect ratio unfriendly for fabrication. In addition, the transformation optics (TO) method is also proposed to transform the straight waveguide in virtual space into a bent waveguide in physical space[9, 20, 21].

It is well known that polarization insensitive devices are favorable to further exploit the bandwidth of the transmission system. However, there are few silicon multimode waveguide bends that can support dual polarizations proposed in the previous works due to the large polarization dependence in most silicon nanophotonic components. Fortunately, the multimode bends with special curves might have much smaller polarization dependence which makes it possible to design polarization insensitive bends with them. In ref[22], a waveguide bend is proposed to support dual polarizations by inducing an Euler-curve assisted with the subwavelength gratings with an equivalent radius of 10μm. This waveguide bend can support six-mode- transmission with low excess losses (<0.23dB) and crosstalks (<-26.5dB) based on the fact that the multimode bends by using the Euler curves can have similar performances for dual polarizations. However, because the grating structure needs to be shallow-etched, second lithography and etching step are needed which increase the fabrication complexity and cost. The excess losses are also slightly raised by the scattering of the gratings, which may become an issue when a large number of bends are cascaded. A review of recent advances in high-performance silicon polarization processing devices also mentions the above device as the only example of the polarization insensitive silicon MWBs[23]. In the previous work we have proposed an ultra-compact silicon MWB structure based on a special curve named as the free-form curve (FFC) [24-25]. Based on this curve a four-mode MWB with the radius of 15μm with ultra-low loss and crosstalk is designed. This FFC-MWB has the advantages of simple fabrication by avoiding the subwavelength gratings and better performance than other special curve-based MWBs. The FFC can be also used for an MWB with dual polarizations. In this paper, we study several polarization-insensitive MWBs based on the FFC, Bezier curve and Euler curve, respectively. The equivalent radius is set to be $R_{eq}$=10μm and the width of the access waveguide is 1.2 μm. The transmission properties of the $TE_0$, $TE_1$, $TE_2$, $TM_0$, $TM_1$, and $TM_2$ mode are simulated and a discussion of the behaviors of these curves is made.

## 2. Structure and Simulation

### 2.1 multimode waveguide bend based on the double free-form curve

It is known from the previous literature that the difference between the refractive indices of the TM and TE modes decreases synchronously with the decrease of the radius. On the contrary, the propagation loss and inter-mode crosstalk both increase when the radius shrinks. Considering the compactness of the MWBs, an equivalent radius of $R_{eq}$=10 μm is selected in this paper. The proposed structure is built in the SOI platform with a 340 nm-thick top Si layer and a 2 μm-thick $SiO_2$ buried layer, and the top cladding material is also chosen to be $SiO_2$. Apparently, a thicker layer (instead of 220 nm) of the Si core helps to reduce the polarization dependence[23]. In addition, this design chooses the width $W$ of the access waveguide to be 1.2 μm to support the transmission of the first three order TE and TM modes with the central operating wavelength of 1550 nm.

In our previous work, a design of a multimode waveguide bend based on the free-form curve is proposed. The FFC is such a smooth curve with inconstant curvature radius of every point on it which can be defined freely. In practice it is set to be self-symmetric and discretized by a series of arcs with different radii for time-saving reason. The MWB based on the single free-form curve (SFFC) has constant width in the bending region, while in this paper we use a modified design where the inner and outer boundaries of the bend are two separate FFCs which makes the waveguide width inconstant in the bending region as shown in Fig. 1. This double free-form curve (DFFC) setup can increase the flexibility of the parameter adjustment and further decrease the bending radius. In the inset of Fig. 1, a detailed description of a FFC with the segment number 2$N$=6 (only show 3 arc segments here due to the symmetry, $N$ is the number of the parameters for one FFC) can be seen. Each adjacent arcs share the same tangent at the connection point so as to avoid additional scattering. In order to minimize the bending radius, the radius of the $i$-th arc $R_i$ should decrease monotonically from the beginning of the curve to the center of symmetry. In the simulation, the discretization number $N$ should be large enough to ensure better results from the optimization, and is chosen as 20 for both the inner and outer FFC considering the performance and simulation time[25]. 3D Finite Difference Time Domain (FDTD) method is used to calculate the crosstalk and excess loss of the proposed MWB. The definition of the figure of merit (FOM) function is related to the crosstalk and excess loss, the FOM function is defined as Eq. (1)[23]:

$$\text{FOM} = 1 - \frac{1}{n}(1-\alpha)\sum_{i}^{n}(1-T_i) - \frac{1}{n}\alpha\sum_{i}^{n}X_i, \qquad (1)$$

$X_i$ and $T_i$ represent the inter-mode crosstalk and transmittance of the $i$-th mode, and $\alpha$ represents the weight factor of the crosstalk in the FOM function. An initial value of $\alpha = 0.5$ is suitable for most situations since both the loss and crosstalk are important to the device's performance.

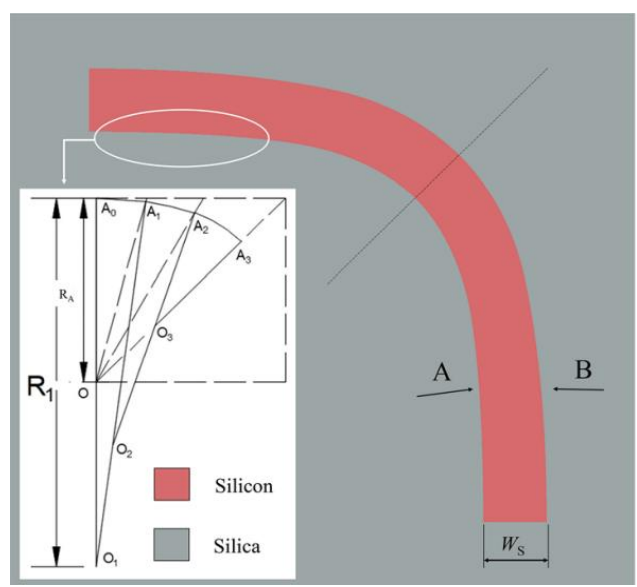

**Fig. 1.** Schematic diagram of the double free-form curves based on the inverse design[24]

The optimization algorithm called direct range search (DRS) is applied here to optimize the combination of a series of arcs with the radii ranging from $R_1$ to $R_N$ [25]. The optimization steps of the DFFC can be briefly summarized as follows.

First, we set a suitable value for the maximum radius of curvature $R_{max}$ (eg $R_{max}=10*R_{eq}$), then an optimization of $R_i$ (*i* from 1 to *N*) is subsequently carried out both for curve A and B. For the search of $R_i$, the value of $R_i$ can be sampled in the range from $R_{i+1}$ to $R_{i-1}$ (from $R_{max}$ to $R_2$ for $R_1$) at equal intervals of 50-100 steps. The value of $R_i$ is kept or discarded depending on the FOM value. All of the arc radii on curve A and B will be optimized in loops until the FOM meets the predefined criterion.

Typically, after tens of hours of optimization by the algorithm, the optimal structure of the DFFC-based MWB can be obtained. The theoretical results of the crosstalk and excess loss in the wavelength range of 1500-1600 nm are calculated, as shown in Fig.2 below.

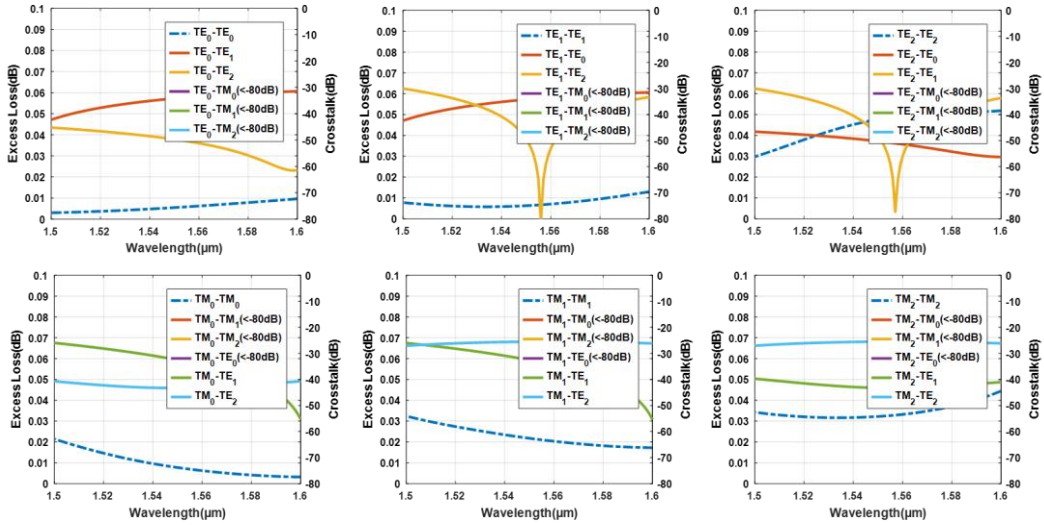

**Fig. 2.** Simulated excess losses and crosstalks of the six modes in the DFFC-MWB.

It can be seen from Fig. 2 that the polarization insensitive MWB with the bending radius of only 10 μm presents very good transmission characteristics of extremely low loss and low crosstalk for all six studied modes. Within the wavelength range of 1500-1600nm, the highest excess loss of each mode is in the range of 0.01-0.052dB, while the worst crosstalk of each mode is in the range of -25.97-31.46dB.

## 2.2 multimode waveguide bend based on the Bezier curve

As we all know that the shape of a cubic Bezier curve is mainly controlled by the parameter *B*, that is, four points (0,0), (R(1-*B*),0), (R, *B*), (R, R) can determine a such curve, as shown in Fig. 3(a). For multimode waveguides, the property of having gradient curvature radius of this curve can just be used to minimize the mode mismatch at the junction of the straight and curved waveguides[26-28]. In order to minimize the crosstalk between the various modes and maximize the transmission efficiency of each mode, it is mainly designed by adjusting the parameter *B* that controls the shape of the curve. Usually, a smaller value of the parameter *B* tends to have larger radius at the starting point of the Bezier curve. This is helpful to reduce the mode mismatch at the junction between the straight and bending waveguide. However, for a fixed equivalent radius, larger radius at the beginning means larger radius gradient along the curve, which increases the radiation loss and crosstalk. When *B*=0.45 the Bezier curve is close to the standard arc[29]. In the following we use the MWB structure composed of two separate Bezier curves as the inner and outer boundaries of the bending part to have better performance, which is different from most of the existed literatures where the two boundaries are independent with constant waveguide width. The main parameters of these two curves are $B_1$ and $B_2$, respectively. When $B_1$ equals to $B_2$, we calculate the insertion loss and crosstalk with different *B* values when $TE_0$ mode is input, as illustrated in Fig. 3(b). The structure tends to have better performance when *B* is within the range of [0.1, 0.4]. Therefore, when optimizing the MWB with two separate Bezier curves as the two boundaries, we set the parameter scanning range to be 0.1-0.4 for $B_1$ and $B_2$.

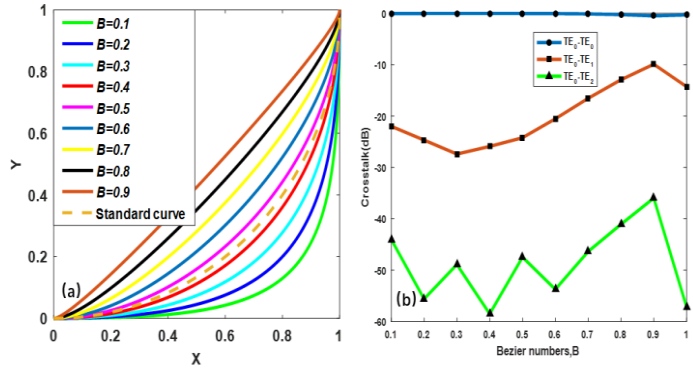

**Fig. 3.** (a) Bezier curves corresponding to different *B* (Bezier numbers, dimensionless). (b) Relationship between the parameter *B* and mode transmission efficiencies (when input mode is $TE_0$ mode).

After tens of hours of calculation with the 3D FDTD method, the results are plotted in Fig. 4. The value of the FOM is plotted with the *B* parameters of the MWB. The definition of the FOM here is the same as for the DFFC-MWB as in Eq. (1). The position with the highest FOM has been marked as the M point in Figure 4, where $B_1$ and $B_2$ are 0.2 and 0.12, respectively. Figure.5 gives the transmission behaviors of all modes in the optimized Bezier-MWB. Within the investigated wavelength range of 1500-1600 nm, both the losses and crosstalks are worse than those of the DFFC-MWB with the same radius. The highest excess loss of each mode is in the range of 0.03-0.18dB, while the worst crosstalk of each mode is in the range of -20.44-27.1dB. Still, the overall performance of the Bezier curve-MWB is also acceptable.

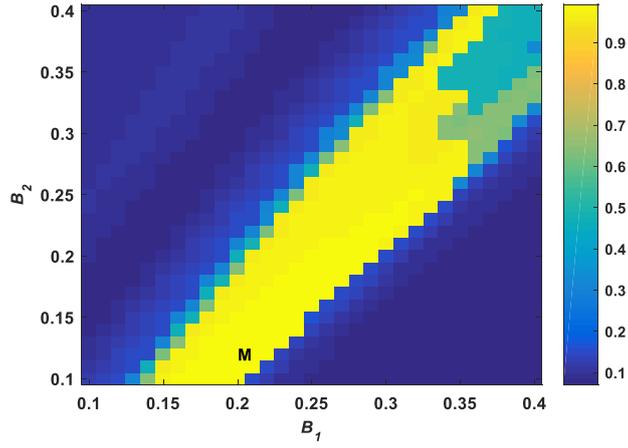

**Fig. 4.** Plotted FOM values with the scanned *B* Parameters of the Bezier curves

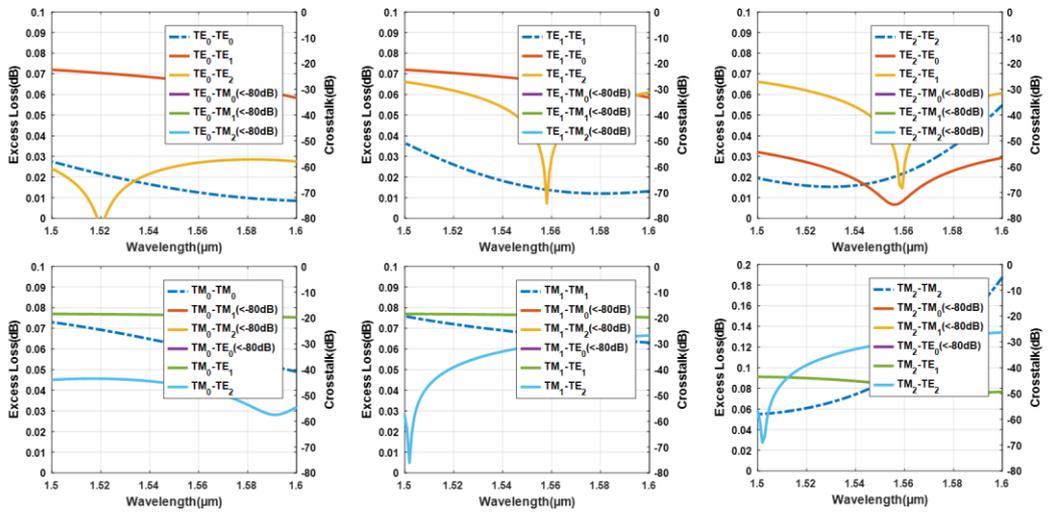

**Fig. 5.** Simulated excess losses and crosstalks of the six modes in the Bezier curve-MWB.

## 2.3 multimode waveguide bend based on the Euler curve

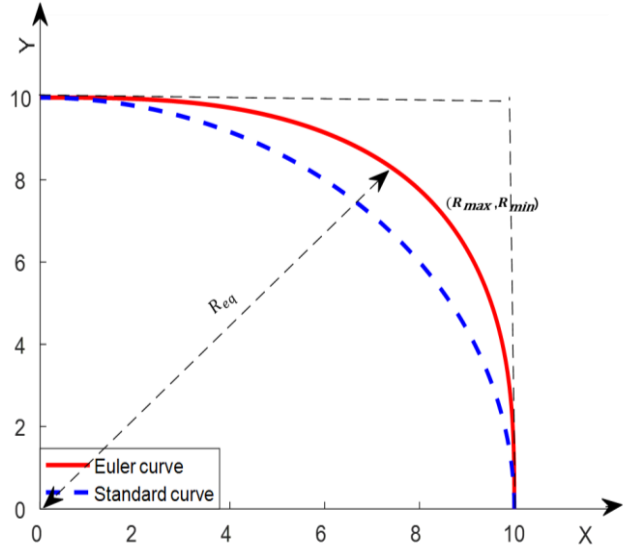

**Fig, 6.** Euler curve and standard curve

The Euler curve has similar property with the Bezier curve in terms of the changing curvature radius along the curve itself and can also be used in the design of waveguide bends[29, 30]. According to the mathematical definition, the curvature of an Euler curve is variable and increases linearly along the curve, as shown in Figure 6. A recent design of silicon MWBs based on the Euler curve can support the simultaneous transmission of 4 TM modes with the equivalent bending radius of 45 μm [14]. The main definition of the curvature radius of the Euler curve can be shown as Eqs (2) and (3). The coordinates of any point on the curve can be derived according to Eqs (4) and (5).

$$\frac{d\theta}{dL} = \frac{1}{R} = \frac{L}{A^2} + \frac{1}{R_{max}}, \tag{2}$$

$$A = [L_0 / (1/R_{min} - 1/R_{max})]^{1/2}, \tag{3}$$

$$x = A\int_0^{L/A} \sin(\frac{\theta^2}{2} + \frac{A\theta}{R_{max}})d\theta, \tag{4}$$

$$y = A\int_0^{L/A} \cos(\frac{\theta^2}{2} + \frac{A\theta}{R_{max}})d\theta, \tag{5}$$

Among them, $R$ is the curvature radius of any point (x,y) on the curve, $L$ is the length of the curve from the starting point (0, 0) to this point (x, y), A is a constant, and $L_0$ is the total length of the whole curve.

In this paper, in order to improve the performance of the MWB as much as possible with the same equivalent radius as the other curves, we also adjust the inner and outer boundaries of the bending waveguide as two independent Euler curves. Similar with the FFC, we set the center of the curve as the symmetric center and the minimal point in terms of the curvature radius. Since the equivalent radius is fixed as 10 μm, the whole curve can be defined when the radius at the starting point is selected. In the previous work, a $R_{max}$ of 600 μm is chosen to reduce the mode mismatch. The $R_{min}$ and $R_{eq}$ are 20 and 45, respectively[14]. In this paper, the equivalent radius is much smaller, a large $R_{max}$ requires a large radius gradient, which may increase the radiation loss and inter-mode crosstalk. Thus, we have sampled different values of $R_{max}$ to estimate a proper range for fine scanning. The results show that the values of 100~300 is reasonable for $R_{max}$. According to this, we calculate the spectral performances of the MWB represented by the FOM function when the initial radius $R_{max}$ of both inner and outer Euler curve are scanned. In Fig. 7 the FOM values are plotted with different $R_{max}$. Here the maximum bending radii corresponding to the inner and outer curve are designated as $R_{max1}$ and $R_{max2}$, respectively. The FOM function is kept the same as Eq. (1), and the marked M point represents the parameters for the maximal FOM value.

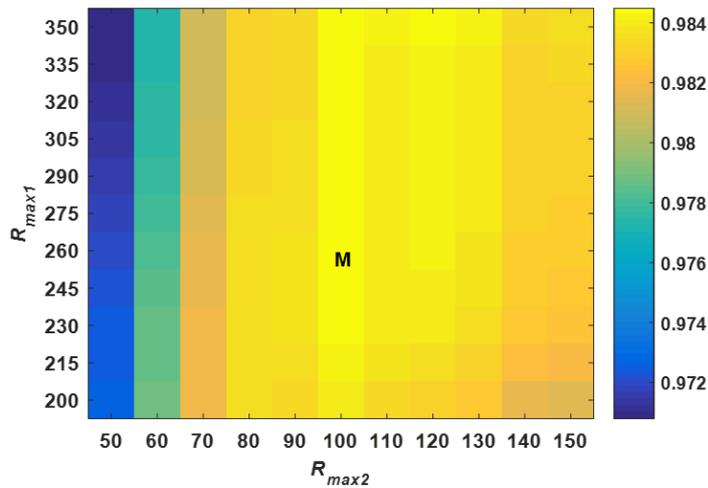

**Fig. 7.** Plotted FOM values with the scanned maximal radii of the Euler curves

After the optimization, the obtained best combination point is marked as M (260,100). Under this condition, minimum radii of the inner and outer curve can be calculated by the above formula to obtain the optimized curves, which can support the transmission of all six modes.

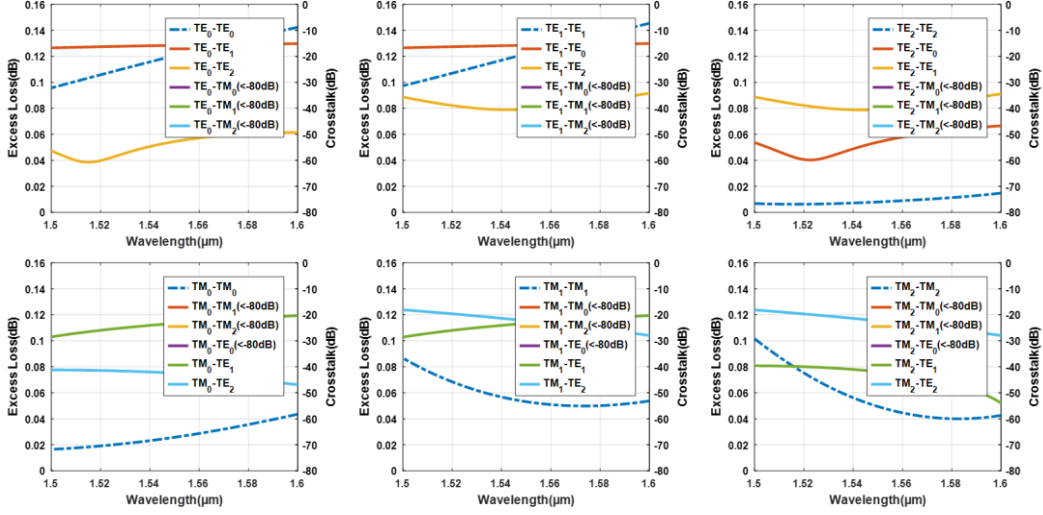

**Fig. 8.** Simulated excess losses and crosstalks of the six modes in the Euler curve-MWB

Fig. 8 gives the excess losses and crosstalks in the wavelength range of 1500-1600nm when the six modes are transmitted, respectively. In the whole band, the highest excess loss of each mode is in the range of 0.02-0.14dB, while the worst crosstalk of each mode is in the range of -15.1-34.4dB. Compared with the other two curves, the average excess loss and crosstalk are both higher. The losses are still lower than those of the SWG-Euler based MWB, but the crosstalks are even exceeding -20dB for some modes. This result implies that the equivalent bending radius is too small for the Euler curve to find a feasible design for highly efficient multimode transmission with the current setup. However, the crosstalk is still much lower than that of a conventional arc-bend waveguide with a radius of 10μm, which is as high as -8.4dB.

**Table 1.**

Excess Loss and Crosstalk of different MWB designs for dual polarizations (λ=1500-1600nm)

| Ref | Optimization Curve | Excess Loss(dB) | Crosstalk(dB) |
|---|---|---|---|
| [22] | Euler with SWGs | <0.23 | <-26.50 |
| This work | Free-form curve | <0.05 | <-25.97 |
| | Bezier curve | <0.18 | <-20.44 |
| | Euler curve | <0.14 | <15.12 |

In Table 1 a comparison of different polarization insensitive MWBs has been made, including the design in the previous work[22]. Apparently, the curve-based MWBs have smaller insertion losses than the design with subwavelength gratings on top. The losses and crosstalks of the MWB based on the Euler curve are obviously higher because this curve fails to achieve high performance with such sharp bending, even when the bend is composed of two separate Euler curves. This is also the reason that the SWGs are applied to assist the highly efficient bending together with the Euler curve in the literature[22]. The double FFC-based MWB has similar crosstalk behavior as the SWG-Euler MWB while much lower loss. Besides, the fabrication step of the FFC-MWB is simple and cost-efficient, and the tolerance is also good according to the experimental results given in reference[25].

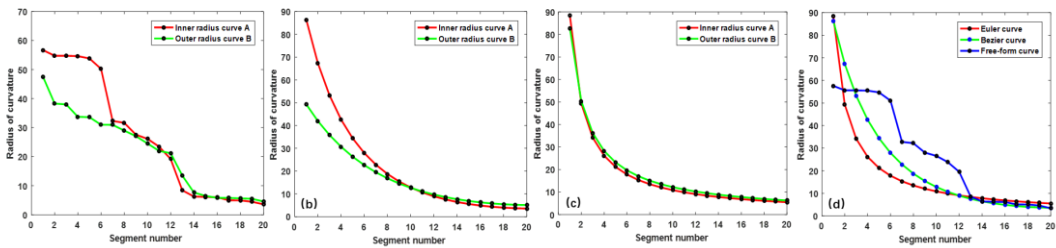

**Fig. 9.** (a), (b), and (c) show the curves of curvature radius corresponding to the inner and outer radii of double free-form curves, Bezier curves, and Euler curves, respectively. (d) gives the comparison of these curves when the initial radii are set similar.

In Fig. 9 we have also shown the curvature distributions of the designed MWBs corresponding to different kinds of curves. In Fig. 9(d), a comparison of the curvature radius distributions in three different types of curves is made when their initial radii are similar. The Bezier curve is one of the boundaries of the optimal bend where the maximal radius is near 90 μm. When $R_{max}$ of the Euler curve is set as the same value, it can be seen that the radius gradient is larger than that of the Bezier curve at the beginning. This may be the reason that the losses and crosstalks of the Euler-MWB are higher, since the radius is changing fast when the straight eigen-modes are coupled into the bent modes. Although the slope variation of the radius values in the Free-from curve is not monotonous, still we can observe that the average gradient for more than half of the segments (from segment 1 to 13 in Fig. 9(d)) is even smaller than that of the Bezier curve. This also implies the benefit of an FFC, where the radius and its gradient can be controlled freely. In theory, no particular mathematics function is capable of giving better performance because the FFC can fit in any specific curve. Thus, we believe the FFC-based MWB for dual polarizations has great potential in the applications of the polarization processing devices.

## 3. Conclusion

In this paper, we have mainly focused on the silicon MWBs suitable for both polarizations based on special curves due to their characteristics of easy fabrication and relatively low loss. MWBs based on the free-form curve, Bezier curve and Euler curve have been simulated and discussed. The theoretical results show that in a very sharp bending of 10 μm, the FFC-based MWB has the optimal transmission properties. The insertion losses of the first three TE and TM modes are extremely small at the level of 0.003~0.05dB in the wide band of 1500-1600nm, and the highest crosstalks of different modes are around -25.97~-31.46dB in the whole wavelength band. The results of this paper provide an efficient method to design the ultra-compact and high performance MWBs for dual polarizations.

# CRediT authorship contribution statement

The manuscript was written through contributions of all authors. All authors have given approval to the final version of the manuscript.

# Data Availability

Data will be made available on request.

# Conflict of interest

The authors declare no conflict of interest.

# Acknowledgments

This work was supported by the Natural Science Foundation of Guangdong Province (no.2020A1515010786) and the Guangdong Rural Science and Technology Commissioner Project (no. ktp20200112).